\documentclass{appolb}
\usepackage{epsfig}
\usepackage{rotating}
\usepackage{graphicx}
\usepackage{bm,amsfonts,amssymb}
\begin{document}
\title{Coupled-channel analysis of the $X(3872)$
}
\author{S. Coito\thanks{Talk at Workshop \em ``Excited QCD 2010'', \em
Stara Lesn\'{a}, Slovakia, 31 Jan.\ -- 6 Feb.\ 2010}, G. Rupp
\address{
Centro de F\'isica das Interac\c c\~oes Fundamentais, Instituto Superior
T\'ecnico, Technical University of Lisbon, P-1049-001 Lisboa, Portugal}
\and
E. van Beveren
\address{
Centro de F\'{\i}sica Computacional, Departamento de F\'{\i}sica,
Universidade de Coimbra, P-3004-516 Coimbra, Portugal}
}

\maketitle

\begin{abstract}
The $X(3872)$ is studied as an axial-vector charmonium state in the
multichannel framework of the Resonance-Spectrum-Expansion 
quark-meson model, previously applied to a variety of other
puzzling mesonic resonances. Included are the open-charm
pseudoscalar-vector and vector-vector channels, the most important of which
is the S-wave $\bar{D}^{*0}D^{0}$+$D^{*0}\bar{D}^{0}$
channel, which practically coincides with the $X(3872)$ structure. The two
free parameters of the model are tuned so as to roughly reproduce the
$\chi_{c1}(3511)$ mass as well as the enhancement just above the
$\bar{D}^{*0}D^{0}$/$D^{*0}\bar{D}^{0}$ threshold. The present model is
able to describe the shape of the latter data quite well. However, as no 
dynamical resonance pole is found, the $X(3872)$ and $X(3940)$ cannot be
reproduced simultaneously, at this stage. A  possible further improvement is
discussed.
\end{abstract}

\section{Introduction}
The $X(3872)$ charmonium-like state was discovered in 2003 by Belle
\cite{PRL91p262001}, as a $\pi^+\pi^- J/\psi$ enhancement
in the process $B^{\pm} \rightarrow K^{\pm} \pi^+\pi^- J/\psi$. The
same enhancement was then seen in several other experiments as well. Finally,
the $X(3872)$ was also observed in the $\bar{D}^0D^0\pi^0$ \cite{PRL97p162002}
and $\bar{D}^{\ast0}D^{0}$ \cite{PRD77p011102} channels. It is now listed
in the PDG tables \cite{PLB667p1}, with a mass of $3872.3\pm0.8$ MeV
and a width of $3.0^{+1.9}_{-1.4}\pm0.9$ MeV.
The observed decay modes and helicity analysis indicate positive $C$-parity
and favour $1^{++}$ or $2^{-+}$ quantum numbers \cite{PLB667p1}.

Theoretical work on the $X(3872)$ comprises a variety of model calculations and
interpretations, because of its seeming incompatibility with the standard
charmonium spectrum. Let us briefly summarise what we consider the most 
significant approaches.
\begin{description}
\item[T\"{o}rnqvist \cite{PLB590p209}]
Calls it a deuson, a deuteron-like $D\bar{D}^*$ bound state. Binding gets
provided by pion exchange. Large isospin breaking is expected, and so 
an important $J/\psi\,\rho$ decay mode \\[-6mm].
\item[Bugg \cite{JPG35p075005}]
Examines a mechanism for resonance capture at thresholds that involves a
threshold cusp interfering constructively with a resonance from
confinement and/or attractive $t$ and $u$-channel exchanges. Conclusion:
the $X(3872)$ requires a virtual state or a resonance \\[-6mm].
\item[Gamermann \& Oset \cite{PRD80p014003}]
Dynamical generation with isospin breaking, in an effective chiral approach. 
Preference for a slightly unbound, virtual
$\bar{D}^{*0}D^{0}$+$D^{*0}\bar{D}^{0}$ state. \\[-6mm]
\item[Lee \textbf{\textit{et al.}} \cite{PRD80p094005}]
Possible $D\bar{D}^*$ molecular bound state in a potential model from
heavy-hadron chiral perturbation theory. \\[-6mm]
\item[Kalashnikova \& Nefediev \cite{PRD80p074004}]
Analysis from data concludes preference for a dynamically generated virtual
$D\bar{D}^*$ state, with a sizable $2\,{}^{3\!}P_1$ $c\bar{c}$ component.
\\[-6mm] 
\item[Danilkin \& Simonov \cite{0907.1088}]
Employs the Weinberg eigenvalue method for a model with 1 confinement and 1
meson-meson channel, coupled via a $^{3\!}P_0$ pair-creation vertex. Two
${}^{3\!}P_1$ peaks are found, i.e., a narrow one at 3.872~GeV and a 
much broader one at 3.935~GeV, a candidate for the $X(3940)$. The $X(3872)$
is of a dynamical nature here. \\[-6mm]
\item[Bugg \cite{JPG37p055002}] 
Discusses several approaches, including the covalent bond \cite{JPG37p055002}.
Proposes resonances as linear combinations of $q\bar{q}$ and meson-meson.
\end{description}
Besides these works, there exists a plethora of exotic models, which 
nevertheless are incapable of describing the $X(3872)$ production data.

\section{Resonance-Spectrum Expansion}
The Resonance-Spectrum-Expansion (RSE) \cite{0809.1149} coupled-channel model
is designed to describe scattering processes of the form $AB \rightarrow CD$.
Although $A, B, C$, and $D$ can in principle be any structures, we usually
consider only non-exotic mesons. Between the initial and final
meson-meson phase, there is an intermediate one that we call the confinement
phase, and it is  built up from an infinite series of $q\bar{q}$ states carrying
the same --- again non-exotic --- quantum numbers.

To describe this picture, we define the effective potential
\begin{equation}
V_{ij}(p_i,p'_j;E)=\lambda^2j^i_{L_i}(p_ia)j^j_{L_j}(p'_ja){\displaystyle
\sum_{n=0}^{\infty} \frac{g_i(n)g_j(n)}{E-E_n}}  \; ,
\label{veff}
\end{equation}
where $p_i$ and $p'_j$ are the relativistically defined relative momenta of
initial channel $i$ and final channel $j$, and $j^i_{L_i}$ is the spherical
Bessel function. As free parameters we have $\lambda$, which is an overall
coupling constant, and $a$, a parameter mimicking the average string-breaking
distance. The relative couplings $g_i(n)$ between meson-meson channels $i$
and $q\bar{q}$ recurrences $n$, with discrete energies $E_n$, are computed
with the formalism developed in Ref.~\cite{ZPC21p291}. The transitions
mechanism is the creation/annihilation of $q\bar{q}$ pairs, according to
the OZI rule, using a spherical Dirac-delta potential, which Fourier
transforms into a spherical Bessel function in momentum space.
Due to the separable form of potential (\ref{veff}), the corresponding
$T$-matrix is obtained in closed form, via the
Lippmann-Schwinger equation with relativistic kinematics \cite{0809.1149}.
Note that the resulting $S$-matrix is manifestly unitary and analytic.

In Eq.~(\ref{veff}), we take a harmonic-oscillator (HO) potential with
constant frequency, not only because of its simplicity but also its success
in many phenomenological applications (see Ref.~\cite{0809.1149} for
references). The corresponding energy levels are \\[-5mm]
\begin{equation}
E_n=m_q+m_{\bar{q}}+\omega(2n+3/2+\ell) \; .
\label{HO}
\end{equation}

\section{$X(3872)$ as an axial-vector charmonium state}
In the present study we assume the $X(3872)$ to be the $2\,{^3\!}P_1$
$c\bar{c}$ state, i.e., the first radial excitation of the $\chi_{c1}(3511)$
\cite{PLB667p1}, with axial-vector ($1^{++}$) quantum numbers. Then, the most
important OZI-allowed decay channels are pseuscalar-vector and vector-vector,
shown in Table \ref{MM}. Note that, for notational simplicity, we have
omitted the bars over the anti-$D$ mesons.
\begin{table}[h]
\centering
\begin{tabular}{|l|c|c|}
\hline
\mbox{}&& \\[-4mm]
channel & relative L & threshold [MeV] \\
\hline 
\mbox{}&& \\[-3.5mm]
$D^0D^{*0}$ & $0$ &$3872$\\ 
$D^{\pm}D^{*\mp}$ & $0$ & $3880$\\
$D_sD_s^*$& $0$ & $4080$\\ 
\hline
\mbox{}&& \\[-3.5mm]
$D^0D^{*0}$ & $2$ &$3872 $\\ 
$D^{\pm}D^{*\mp}$ & $2$ & $3880 $\\
$D_sD_s^{*}$& $2$ & $4080$\\
\hline 
\mbox{}&& \\[-3.5mm]
$D^{*}D^{*}$& $2$ & $4018$\\ 
\hline
\end{tabular}
\mbox{} \\[1.5mm]
\caption{Included meson-meson channels}
\label{MM}
\end{table}

\section{Results and conclusions}
Before studying resonance poles in the vicinity of the $X(3872)$, we must
first fix the model parameters. For the charm quark mass $m_c$ and the 
universal HO frequency $\omega$, we take the values 1.562~GeV and 0.19~GeV,
respectively, which have been kept unaltered in all previous work, starting
in Ref.~\cite{PRD27p1527}. As for $a$, one of the two free
parameters, we take 2.0~GeV$^{-1}$ (about 0.4~fm), which is roughly in
agreement with the value chosen in
Ref.~\cite{PRD80p094011} for $s\bar{s}$ vector resonances, scaled with an
expected factor of $\sqrt{m_s/m_c}$. The overall coupling $\lambda$
we use here as a tunable parameter in order to study the behaviour of 
complex-energy poles, which can be found numerically with ease, since 
the $T$-matrix is known analytically. An indication for the approximate
value of $\lambda$ may come from the mass of the $\chi_{c1}(1P)$
\cite{PLB667p1}, viz.\ $3511$~MeV.

Table~\ref{poles} shows some pole positions for different
values of $\lambda$.
\begin{table}[ht]
\centering
\begin{tabular}{|c|c|c|}
\hline
\mbox{}&& \\[-3.5mm]
$\lambda$ [GeV$^{-3/2}$] & Pole 1 [MeV] & Pole 2 [MeV]\\ 
\hline
\mbox{}&& \\[-3.5mm]
$3.2$ & $3551$ &$3871$\\ 
$3.0$ & $3555$ &$3873-i1$\\ 
$2.2$ & $3572$ & $3884-i6$\\
$1.2$ & $3590$ & $3928-i16$\\ 
\hline
\end{tabular}
\mbox{} \\[1.5mm]
\caption{Pole positions as a function of $\lambda$}
\label{poles}
\end{table}
The poles labelled 1 and 2 originate from the HO confinement states at
$E_0=3599$~MeV and $E_1=3979$~MeV (cf. Eq.~(\ref{HO}) with $\ell=1$),
respectively, for $\lambda\!\to\!0$. These we call confinement poles.
We also searched for dynamical --- or continuum --- poles, which cannot be
simply linked to a confinement state and, moreover, disappear in the
continuum for $\lambda\!\to\!0$, with an ever increasing negative
imaginary part.
Such poles we observed e.g.\ in Ref.~\cite{PRD80p094011}, though with a large
width. Here, if one showed up close to the lowest threshold, just as in
Ref.~\cite{0907.1088}, it might explain the $X(3872)$ structure as a dynamical
resonance, but none was found. We shall come back to this point below. 
\begin{table}[b]
\centering
\mbox{} \\[-2mm]
\begin{tabular}{|c|c|c|c|c|}
\hline
\mbox{} \\[-4mm]
& Re $E$ & Im $k$ & Re $k$ &type of pole\\ 
\hline
\mbox{} \\[-4mm]
1 & $> Th$ & $< 0$ & $> 0$ & resonance\\ 
2 & $< Th$ & $< 0$ & $> 0$ & virtual resonance\\ 
3 & $< Th$ & $< 0$ & $= 0$ & virtual bound state\\
4 & $< Th$ & $> 0$ & $= 0$ & bound state\\ 
\hline
\end{tabular}
\mbox{} \\[1.5mm]
\caption{Pole 2 in different Riemann sheets (see text and Fig.~1)}
\label{rsheets}
\end{table}

Focusing now on Pole 2, which comes out close to the $X(3872)$ structure for
$\lambda\sim3.0$~GeV$^{-3/2}$, we plot in Fig.~1 its trajectory 
around the
\parbox[t]{90mm}{\hspace*{-1.4cm}
\epsfxsize=90mm
\epsfbox{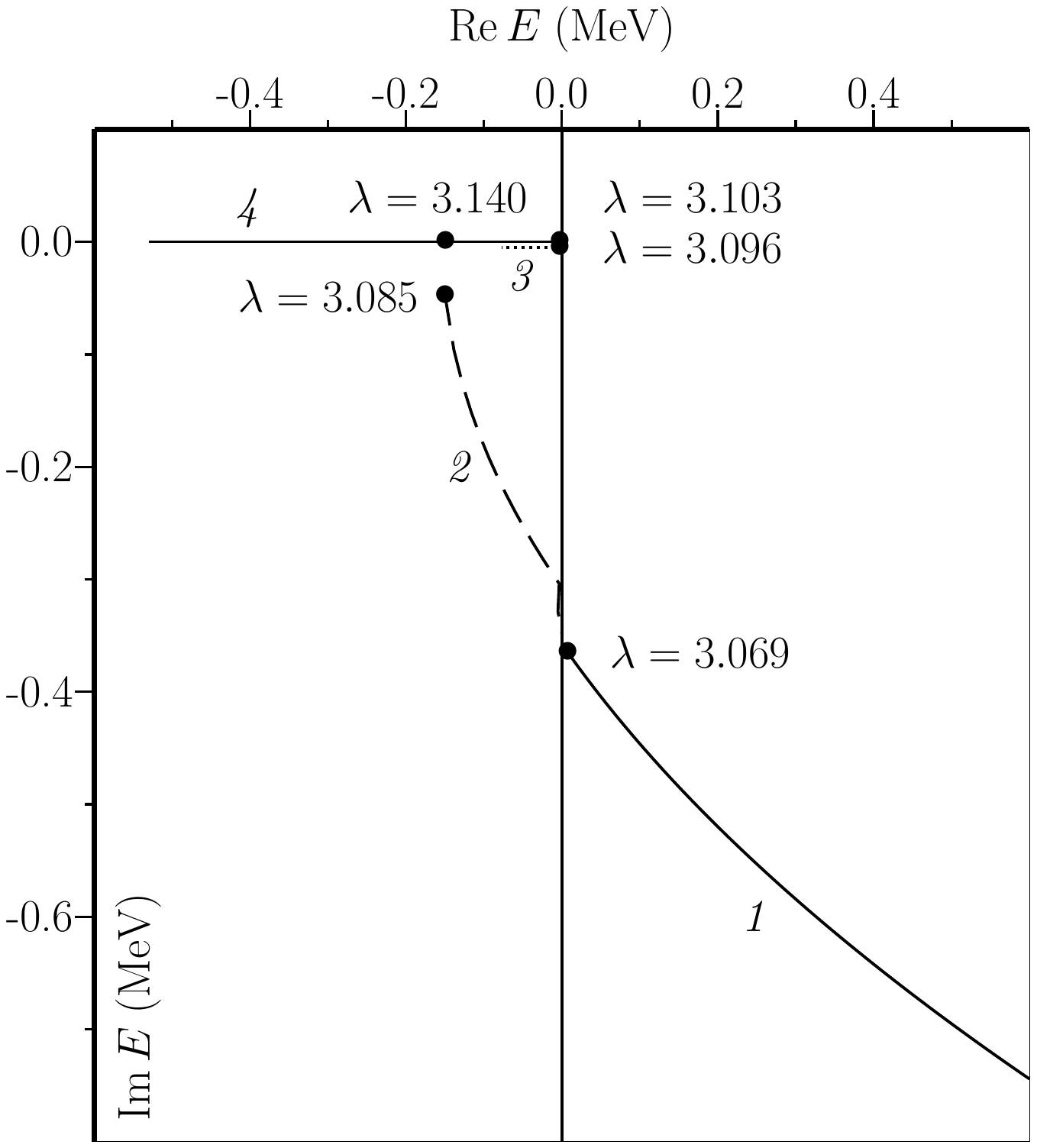} \\[-4.9cm] \small\baselineskip 3.5mm
Fig. 1.  Trajectory of Pole 2: from a \\ resonance to a bound state.
} \hspace*{-2.6cm} \parbox[t]{60mm}{\mbox{}\\[-114mm]
 $\bar{D}^{*0}D^{0}$/$D^{*0}\bar{D}^{0}$ threshold, through two
different Riemann sheets (RSs). The trajectory can be split into four
branches according to the classification in Table \ref{rsheets}, where
$Th$ stands for the threshold energy. We change
from one RS to another by choosing a specific root of the relative
momentum $k$. The first three branches lie on one  RS, and the fourth 
on another RS. We denote branch 2 by \em ``virtual resonance'', \em as
it concerns a resonance with negative phase space. This is a phenome\-non
unique to $S$-wave thresholds. For other waves, a resonance pole always
\hfill turns \hfill into \hfill a \hfill virtual \hfill bound} \\[-3.75cm]
state directly, without the virtual-resonance branch.

\setcounter{figure}{1}
In Fig.~\ref{Tsqp1}, left-hand graph, we depict the elastic $D^0D^{\ast0}$
amplitude just above threshold, for 4 values of $\lambda$ corresponding
to the 4 situations described in Table~\ref{rsheets} and in Fig.~1. As
a matter of fact, we plot the quantity $k|T|^2$, which is the one to
be compared \cite{PRD80p014003} with experiment. The scarce data, not
shown here, cannot really distinguish between the 4 scenarios described
above, though we conclude that a bound state is less likely
\cite{preparation}. \\[-7mm]
\begin{figure}[hb]
\hspace*{-35pt}
\begin{tabular}{lr}
\resizebox{!}{330pt}{\includegraphics{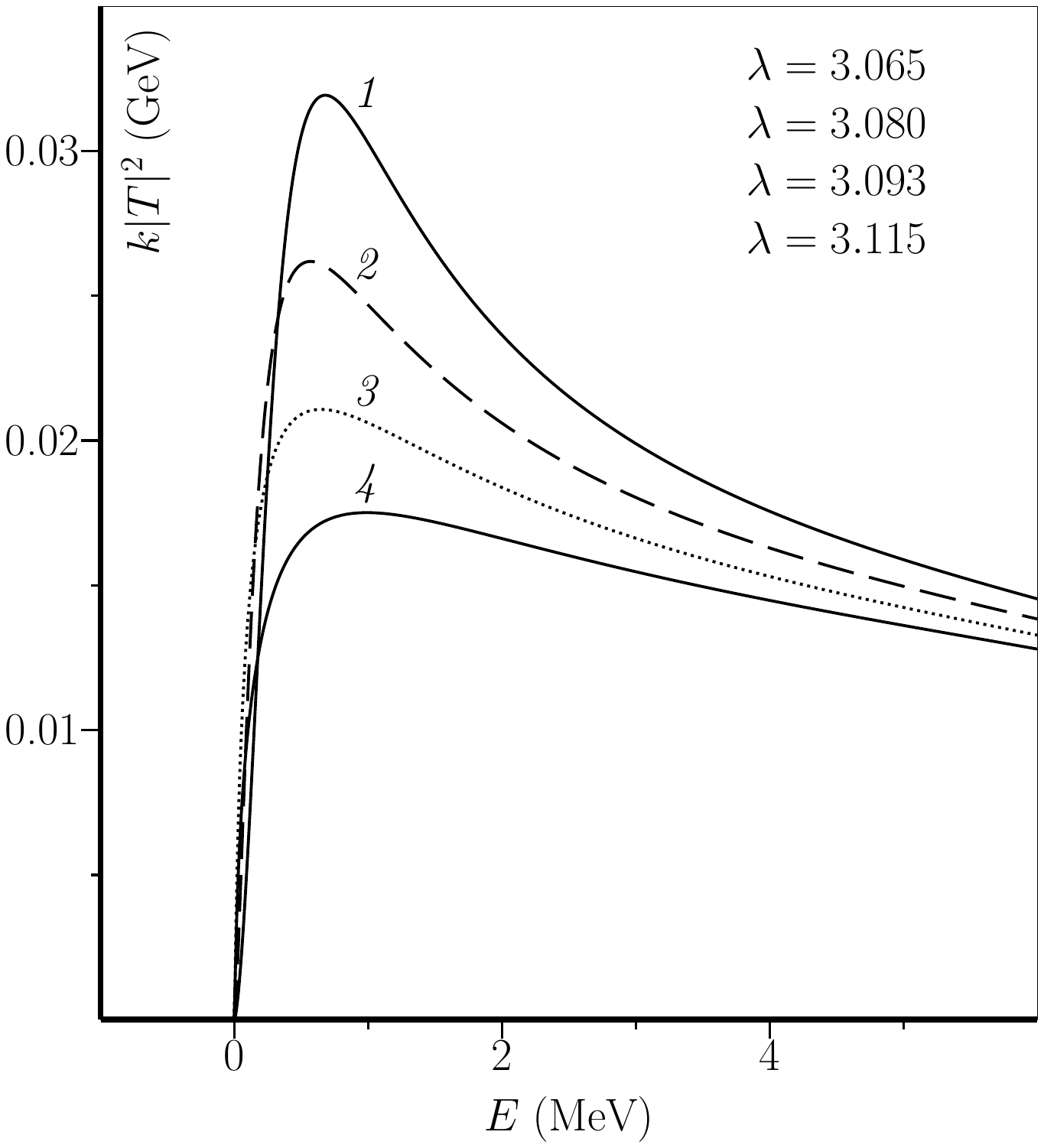}}
&
\hspace*{-85pt}\resizebox{!}{330pt}{\includegraphics{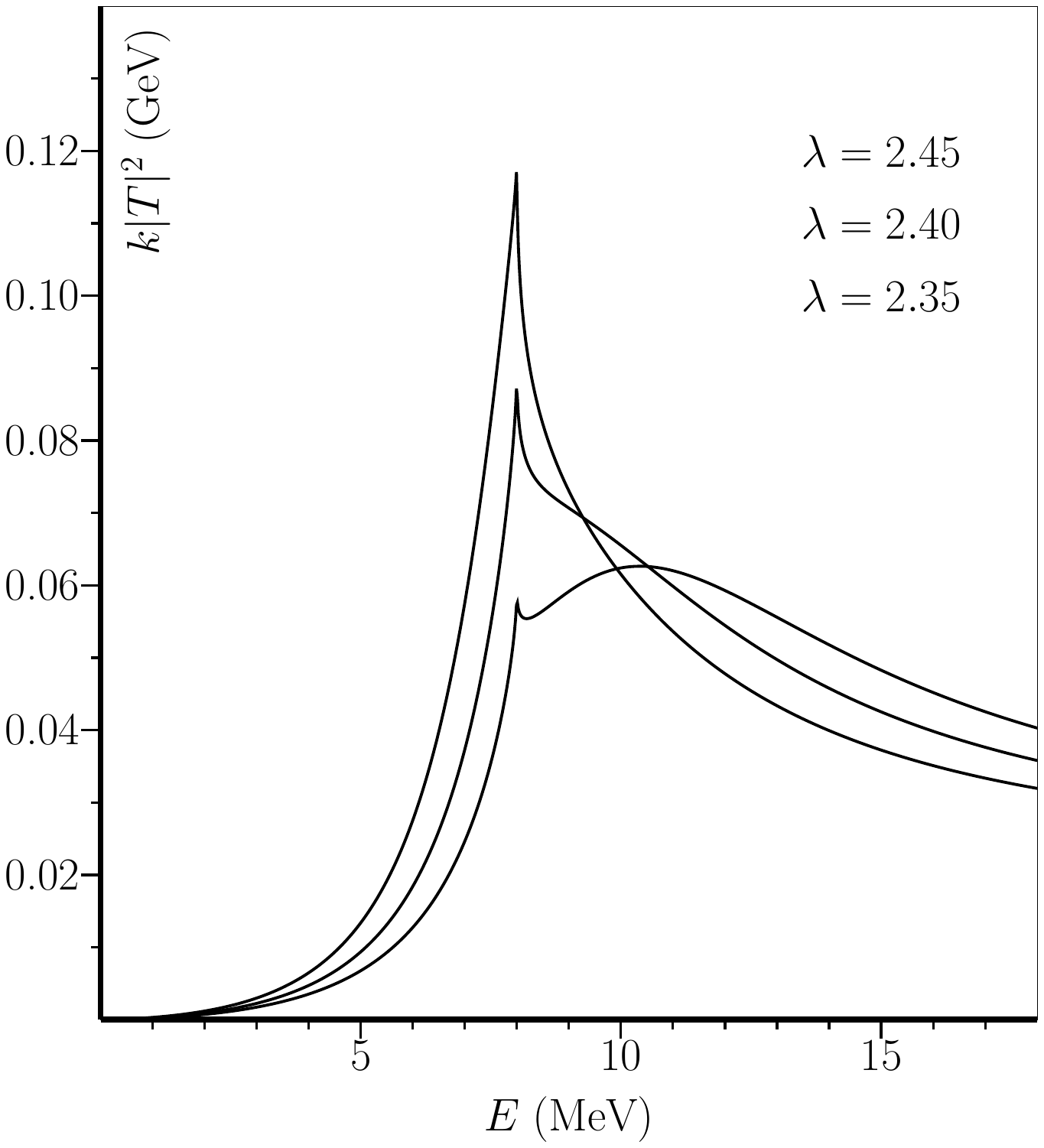}}
\end{tabular}
\mbox{} \\[-5.7cm]
\caption{$k|T|^2$ for elastic $D^0D^{\ast0}$ scattering
(0 MeV corresponds to 3872 MeV).}
\label{Tsqp1}
\end{figure}

Finally, in Fig.~\ref{Tsqp1}, right-hand graph, we generate a cusp structure
around the $D^\pm D^{\ast\mp}$ threshold, by reducing $\lambda$ and so moving
Pole 2 upwards. While quite illustrative, such a scenario is now excluded by
the data.

In conclusion, we have studied the $X(3872)$ enhancement as the
$2\,{}^{3\!}P_1$ $c\bar{c}$ state in the RSE coupled-channel model. 
The included OZI-allowed channels are capable of reproducing the observed
structure in $D^0D^{\ast0}$, via a nearby (virtual) resonance or
virtual-bound-state pole. However, the present model does not generate
any dynamical pole close to the real axis, so that the $X(3940)$ 
\cite{PLB667p1} --- if
indeed a $1^{++}$ state as the $X(3872)$ --- remains unexplained. We are
now working \cite{preparation} on model extensions, e.g.\ by including the
probably observed \cite{PRL96p102002}, OZI-violating $J/\psi\,\rho$ decay mode.

We thank the organisers for another great workshop. We are
also indebted to D.~V.~Bugg for very useful discussions on
the nature of the $X(3872)$.
This work was supported in part by
the \emph{Funda\c{c}\~{a}o para a Ci\^{e}ncia e a Tecnologia}
\/of the \emph{Minist\'{e}rio da Ci\^{e}ncia, Tecnologia e Ensino Superior}
\/of Portugal, under contract CERN/FP/109307/2009, and by Instituto Superior
T\'{e}cnico, through fellowship no.\ SFA-2-91/CFIF.

\end{document}